  \documentclass{aa}  

  \def\xmm{{\it XMM-Newton~}} 
  \def\swift{{\it Niel Gehrels Swift~}} 
  \def\chandra{{\it Chandra~}} 
  \def\nustar{{\it NuSTAR~}}

  \usepackage{natbib}
  \usepackage{graphicx}
  \usepackage{longtable}
  \usepackage{amsmath}
  \usepackage{calrsfs}
  \usepackage{physics}

\DeclareMathAlphabet{\pazocal}{OMS}{zplm}{m}{n}%

\newcommand{\La}{\mathcal{L}}%
\newcommand{\LX}{L_\mathrm{X}}
\newcommand{\mnh}{N_\mathrm{H}}

\newcommand{\dlog}{\mathrm{dlog}}

  \usepackage{txfonts}

\begin{document}

     
      \titlerunning{Compton-thick AGN}
      \authorrunning{I. Georgantopoulos et al.}
      
      \title{The Compton-thick AGN luminosity function in the local Universe: }
      \subtitle{A robust estimate combining BAT detections and \nustar spectra}

     \author{I. Georgantopoulos \inst{1}, E. Pouliasis \inst{1}, A. Ruiz\inst{1}, A. Akylas\inst{1}}    
     \institute{Institute for Astronomy Astrophysics Space Applications and Remote Sensing (IAASARS),\\ National Observatory of Athens, \\ Ioannou Metaxa \& Vasileos Pavlou, Penteli, 15236, Greece\\
     \email{ig@noa.gr}
      }


    \abstract{The Compton-thick Active Galactic Nuclei (AGN) arguably constitute the most elusive class of sources as they are absorbed by large column densities above $\rm  10^{24} cm^{-2}$. These extreme absorptions hamper the detection of the central source even in hard X-ray energies. In this work, we use both \swift and \nustar observations in order to derive the most accurate yet Compton-thick AGN luminosity function. We, first, compile a sample of candidate Compton-thick AGN ($\rm N_H=10^{24-25}~cm^{-2}$) detected in the Swift BAT all-sky survey in the 14-195 keV band. We confirm that they are Compton-thick sources by using the follow-up \nustar observations already presented in the literature. Our sample is composed of 44 sources, consistent with a column density of $\rm 10^{24}-10^{25}~cm^{-2}$ at the 90\% confidence level. These have intrinsic luminosities higher than $\rm L_{10-50keV} \sim 3\times10^{41}~erg~s^{-1}$ and are found up to a redshift of z=0.05 ($\approx$200 Mpc). We derive the luminosity function of Compton-thick AGN using a Bayesian methodology where both the full column density and the luminosity distributions are taken into account. The faint end of the luminosity function is flat, having a slope of $\gamma_1=0.01^{+0.51}_{-0.74}$, rather arguing against a numerous population of low luminosity Compton-thick AGN. Based on our luminosity function, we estimate that the fraction of Compton-thick AGN relative to the total number of AGN is of the order of 24$\pm$5\% in agreement with previous estimates in the local Universe based on BAT samples. 
    }


    \authorrunning{Georgantopoulos et al.}

    \maketitle
  %

	\section{Introduction}
Active Galactic Nuclei (AGN) are among the most luminous sources in the Universe. They are powered by accretion onto supermassive black holes (SMBHs) in their centres \citep{Lynden1969}. As X-ray emission is ubiquitous in AGN, the \xmm and \chandra X-ray missions have probed with unprecedented accuracy the AGN demographics and hence they have mapped in detail the accretion history in the Universe \citep[e.g.][]{Ueda2014, Aird2015, Miyaji2015, Ranalli2016, Fotopoulou2016, Georgakakis2017, Pouliasis2024}. The majority of AGN are obscured by large amounts of dust and gas. This obscuring screen is believed to have the form of a torus, although direct imaging at mid-IR and sub-mm wavelengths reveals a more complicated structure \cite[e.g.][]{Honig2012,Garcia-Burillo2016}.

Additional constraints on the demographics of the AGN population come from the diffuse X-ray radiation that permeates the whole Universe, the X-ray background \citep{Giacconi1962}. This radiation comes from the superposition of all discrete point sources in their vast majority AGN \citep{Mushotzky2000}. The energy density of the X-ray background peaks at around 30 keV \citep{Frontera2007, churazov2007, revnivtsev2003}. The population synthesis models that attempt to reproduce the shape of  the X-ray background, and especially its 30 keV hump, 
 find that a significant fraction of the sources that constitute this radiation must be associated with Compton-thick sources \citep{Comastri1995,Gilli2007,Treister2009,Akylas2012,Ananna2019}.  
These heavily obscured sources present column densities in excess of $\approx 10^{24}$ $\rm cm^{-2}$ and hence the obscuration is caused by Compton scattering on electrons rather than photoelectric absorption \citep[e.g.][]{Hickox2017}. Nevertheless, the X-ray background synthesis models appear to produce divergent results regarding the number of Compton-thick sources. On the low end of the models of \cite{Akylas2012} and \cite{Treister2009} yield Compton-thick fractions of less than 20\%. On the other end the models of \cite{Ananna2019} predict Compton-thick fractions of the order of 50\%. Recently, \citet{Carroll2023} attempted to constrain the fraction of Compton-thick sources by forward modelling the column density distribution of mid-IR selected AGN up to a redshift of $z=0.8$. They estimate a Compton-thick fraction of 55\% \cite[but see][]{Georgakakis2020}.

A number of works attempted to detect directly these highly obscured sources using the superb sensitivity of the \xmm and \chandra missions \citep{Georgantopoulos2009,Georgantopoulos2013,Lanzuisi2015,Brightman2014, Koulouridis2016, Buchner2014,Lanzuisi2017, Georgakakis2017, Corral2019, Laloux2023}. The identification of the Compton-thick sources is based on either the direct detection of the absorption turnover or  the detection of a large equivalent width FeK$\alpha$ line which is the smoking gun of heavy obscuration. However, as these missions operate in the relatively soft 0.5-10 keV band, they may miss a significant number of these heavily obscured sources.

Instead, the BAT (Burst Alert Telescope) instrument \citep{Barthelmy2005}, on board the \swift mission, \citep{gehrels2004}, all-sky survey was very prolific in detecting large number of candidate Compton-thick AGN owing to its energy passband which extends to very hard energies $\rm 14-195~keV$. The 70-month BAT survey detected 728 non-blazar AGN. \cite{Ricci2015} identified 55 Compton-thick AGN among this sample up to a redshift of $z\sim0.1$. Using again the 70-month BAT sample, \cite{Akylas2016} applying a Bayesian approach, identified a few tens of candidate Compton-thick AGN assigning a column density probability distribution to each object. Both \cite{Ricci2015} and \cite{Akylas2016} found an {\it observed} Compton-thick fraction of less than 10\% of the total AGN population. However, when \cite{Torres-Alba2021} used a volume-limited sample of Compton-thick AGN with \nustar spectra up to $z=0.01$, they estimated an {\it intrinsic} Compton-thick fraction of about 20\%. Their results refer  to the column density range $\rm N_H=10^{24-25}~cm^{-2}$, so called transmission dominated Compton-thick sources. This is the same column density range routinely probed by the 
 \cite{Ricci2015} and \cite{Akylas2016} works. The work of \cite{Torres-Alba2021} suggests that a number of Compton-thick sources are so heavily obscured that their detection is arduous even in the BAT high energy band \cite[see also][]{Burlon2011}.  Some X-ray background synthesis models \citep[e.g.][]{Ananna2019} predict much higher Compton-thick fractions up to 50\%. This is because 
 they assume an 
equally numerous fraction of Compton-thick AGN with extreme column densities in the range $\rm 10^{25-26}~cm^{-2}$, dubbed as reflection-dominated sources.
 
Based on their BAT sample \cite{Akylas2016} derived the luminosity function of Compton-thick AGN. \cite{Ananna2022} derived the luminosity function using the \cite{Ricci2015} Compton-thick BAT sample finding a reasonable agreement with the work of \cite{Akylas2016}. However, it soon became evident that the BAT's moderate sensitivity and spectral resolution may hamper the reliable identification of a number of faint sources as bona-fide Compton-thick AGN. For this reason \nustar observations of the BAT selected AGN have been systematically employed in order to pin-point accurately the exact absorbing column density \citep[e.g.][]{Tanimoto2022,Torres-Alba2021,Marchesi2017, Marchesi2018, Georgantopoulos2019,Zhao2021,Silver2022}. 

Here, we compile a local, $z<0.05$, sample of BAT selected candidate Compton-thick AGN. All these have \nustar observations available. The secure column density determinations by \nustar allow us to revisit the Compton-thick AGN luminosity function and its absorption correction in the local Universe. Using this luminosity function we provide a robust estimate of the fraction of Compton-thick AGN in the local Universe. Throughout the paper, we assume a standard $\rm \Lambda CDM$ cosmology with $\rm H_o=69.3km s^{-1}Mpc^{-1}$, $\rm \Omega_m=0.286$.

 \begin{figure}
\center
\begin{tabular}{c}
    \includegraphics[width=0.47\textwidth]{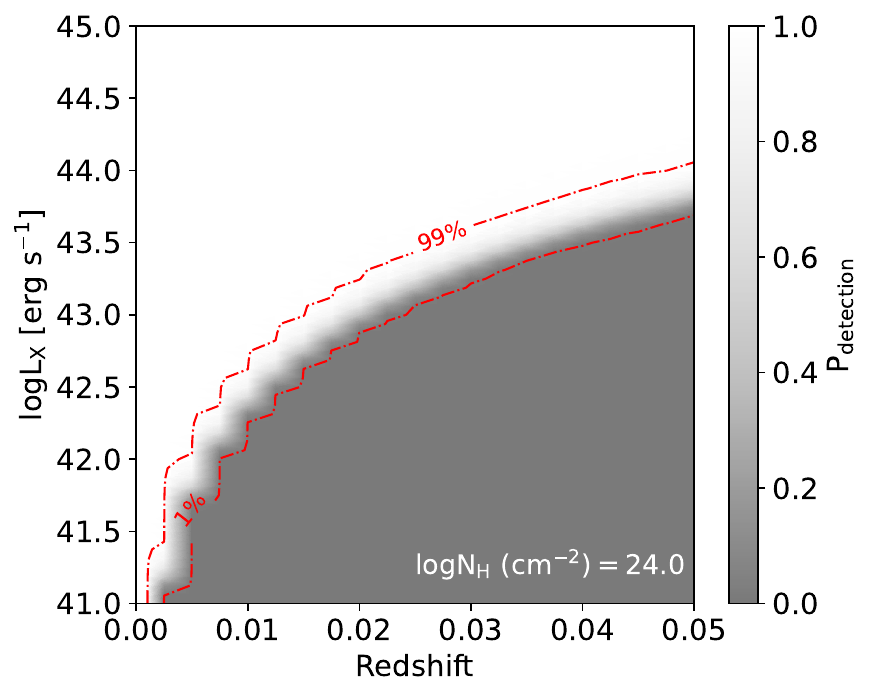} \\
    \includegraphics[width=0.47\textwidth]{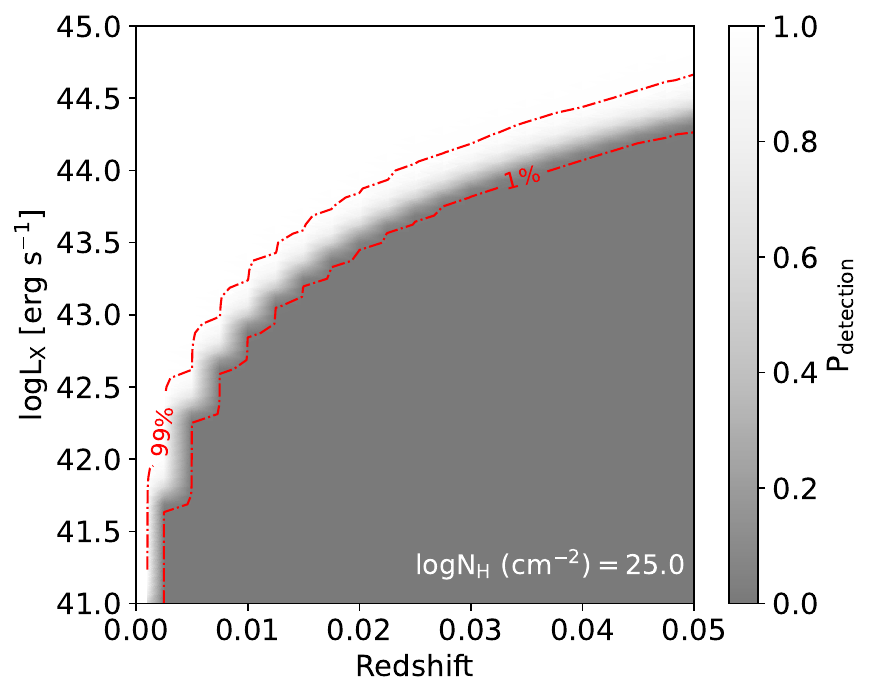} 
\end{tabular}
\caption{Detection probability maps as a function of the intrinsic X-ray luminosity and redshift for a source with intrinsic column density of $\rm N_H \sim 10^{24}~cm^{-2}$ (upper) and $\rm N_H \sim 10^{25}~cm^{-2}$ (lower). The red lines indicate the probability of detection at 1\% and 99\%.}\label{senmap}
\end{figure}

\begin{figure}
\centering
\includegraphics[width=0.4\textwidth]{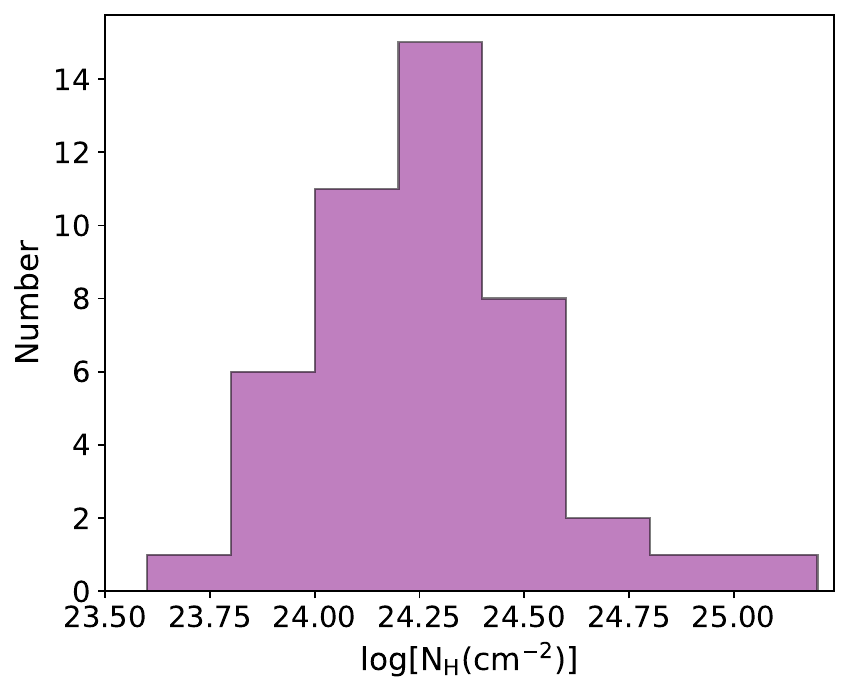}\\
\includegraphics[width=0.4\textwidth]{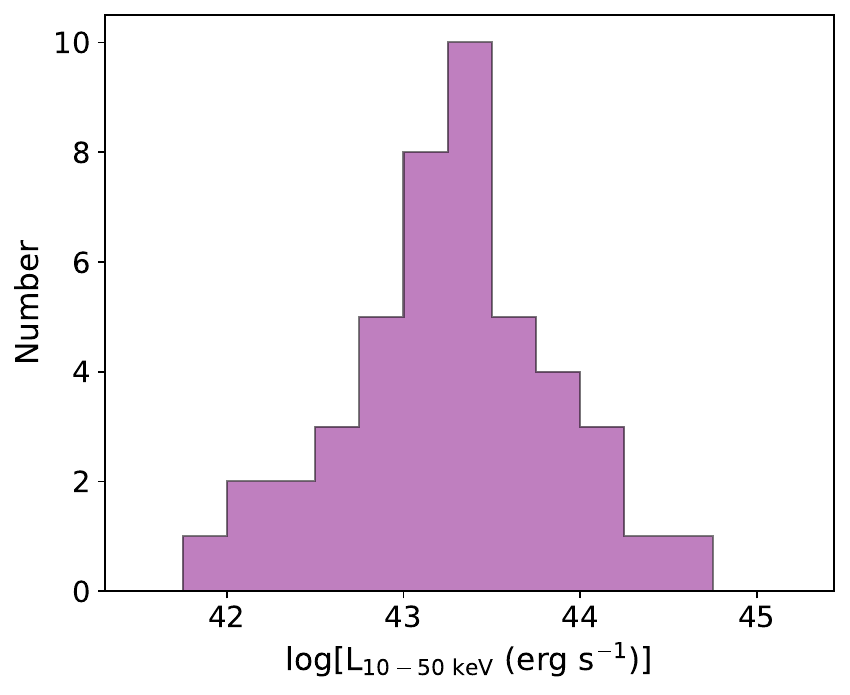}\\
\includegraphics[width=0.4\textwidth]{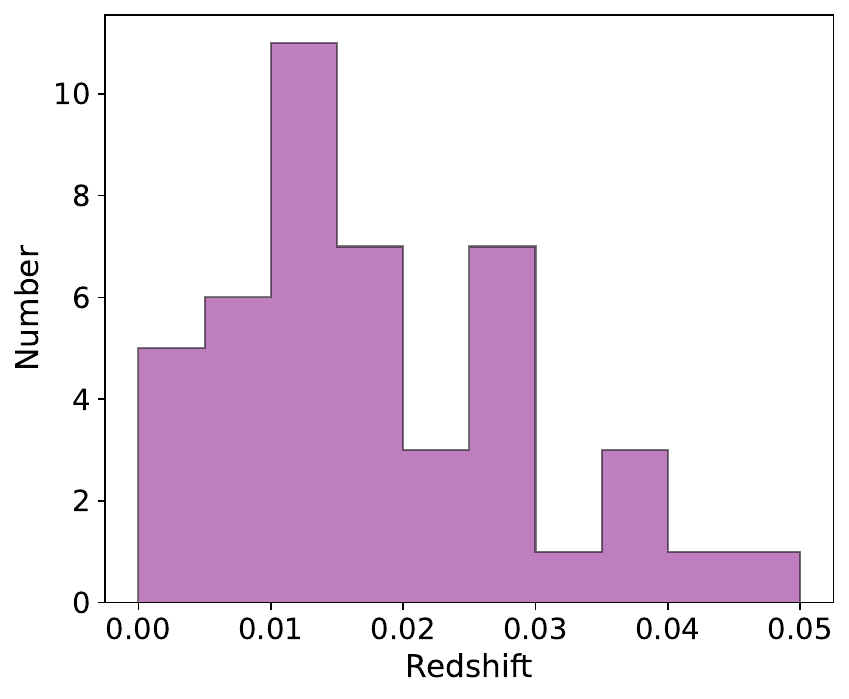}
\caption{Distributions of the Hydrogen column density (top panel), the $10-50$~keV absorption corrected luminosity (middle panel) and redshift (bottom panel) for our sample.}
\label{distributions}
\end{figure}


\begin{figure}
\centering
\begin{tabular}{c} 
\includegraphics[width=0.40\textwidth]{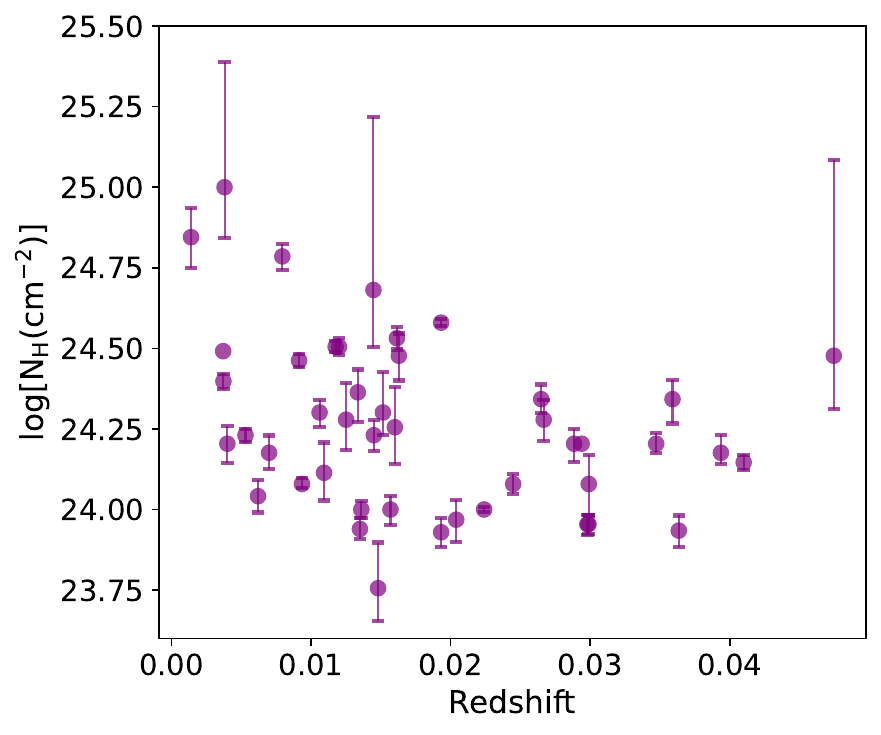}
\\
    \includegraphics[width=0.40\textwidth]{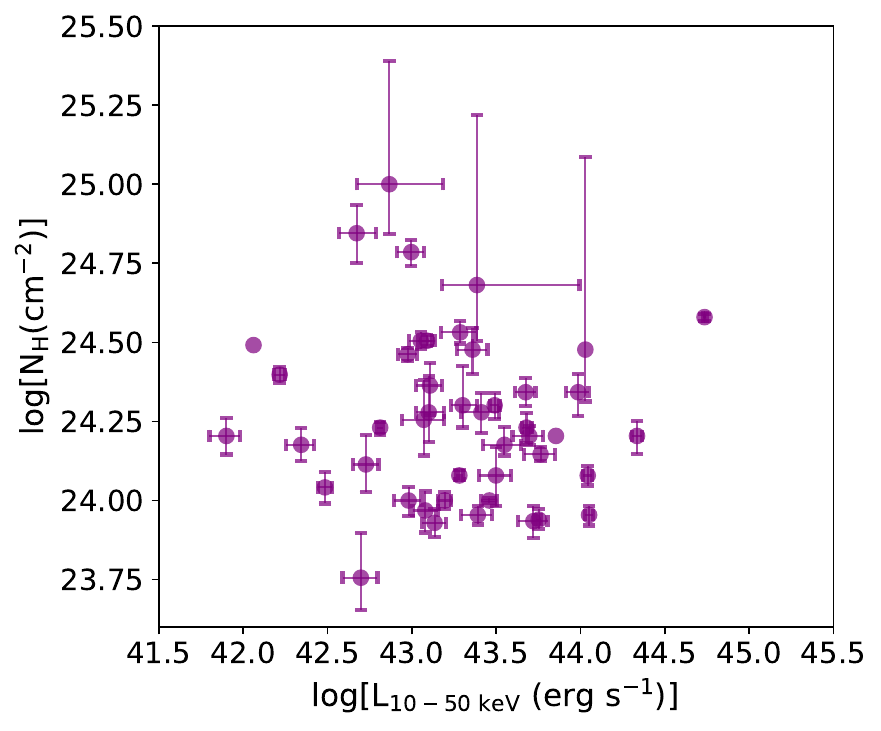} 
\\
    \includegraphics[width=0.40\textwidth]{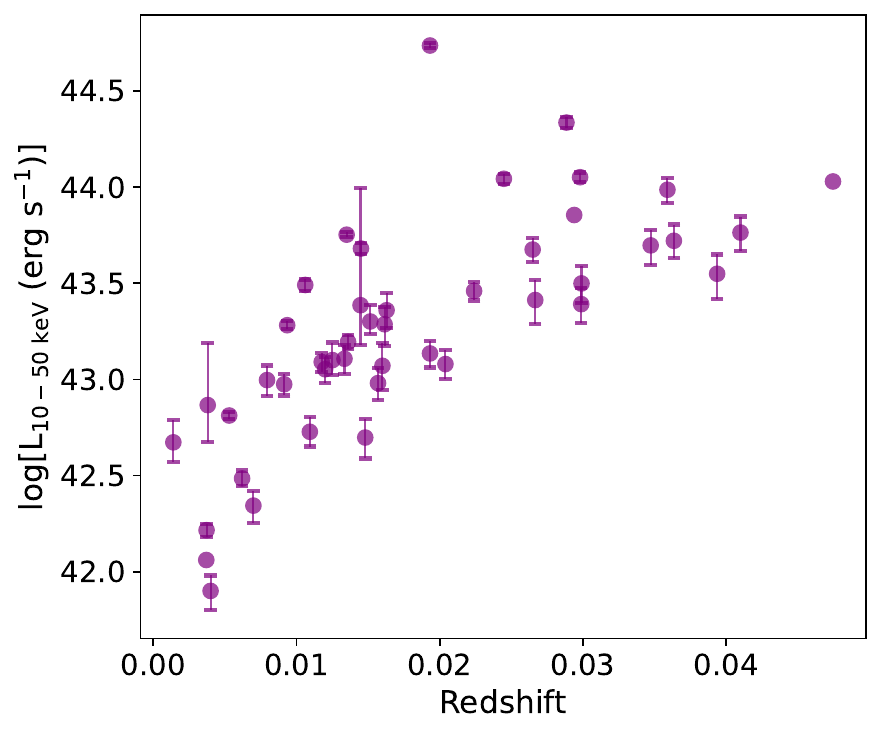}    
\end{tabular}
\caption{Column density vs. redshift (top panel); 
column density vs. X-ray absorption corrected luminosity (middle panel); 
X-ray absorption-corrected, luminosity vs. redshift (bottom panel).  }
\label{z-lx}
\end{figure}

\section{Data}

\subsection{\swift}
	The {\it Swift} Gamma-Ray Burst (GRB) observatory \citep{gehrels2004} was launched in November 2004 and has been
	continually observing the hard X-ray ($\rm 14-195$ keV) sky with the Burst Alert Telescope (BAT). BAT is a large, coded-mask telescope, optimized to detect transient GRBs and it is designed with a very wide field-of-view of $\sim$ 60$\times$100 degrees.

	The data presented in this paper stem from the analysis of the sources detected in the 70 months of observations 
	of the BAT hard X-ray detector on the \swift Gamma-Ray Burst observatory \citep{baumgartner2013}. The 70-month BAT survey is an almost uniform, hard X-ray, all-sky survey with a sensitivity of 1.34 $\times10^{-11}$ ergs s$^{-1}$ cm$^{-2}$ over 90\% of the sky and 1.03$\rm \times10^{-11}$ ergs s$^{-1}$ cm$^{-2}$ over 50\% of the sky, in the 14-195 keV band. 
	
  \subsection{NuSTAR}
	The Nuclear Spectroscopic Telescope Array \citep[\nustar,][]{harrison2013}, launched in June 2012, is the first
	orbiting X-ray observatory which focuses light at high energies (E $>$ 10 keV). It consists of two co-aligned focal plane modules (FPMs), which are identical in design. Each FPM covers the same 12 x 12 arcmin portion of the sky, and comprises of four Cadmium-Zinc-Tellurium detectors. \nustar operates between 3 and 79 keV, and
	provides an improvement of at least two orders of magnitude in sensitivity compared to previous hard X-ray observatories operating at energies E$>$10 keV. This is because of its high spatial resolution 58 arcsec Half-Power-Diameter.

  \subsection{The Compton-thick AGN in the BAT survey}\label{BAT_CT}
  1171 hard X-ray sources have been detected in the 70-month BAT survey down to 4.8$\sigma$ associated with 1210 counterparts. The majority of these sources are associated with AGN. In particular, there are 752 non-blazar AGN among these sources \citep{Koss2022}. \cite{Ricci2015} derived the X-ray spectra for the 70-month BAT survey sources combining the \swift spectra with available \xmm, \chandra {\it SUZAKU} and {\it ASCA} data. The X-ray spectra showed that 55 sources are associated with Compton-thick obscuration. In another study \cite{Akylas2016} using again the BAT 70-month survey find 42 sources with a probability above 30\% of being Compton-thick. Although there is a significant overlap between the two samples there is a number of sources which have been characterised as Compton-thick in one sample but not in the other. Specifically, 35 sources are common between the two samples. Apart from the different methodology, it has to be noted that \cite{Akylas2016} used only the \swift (BAT and the X-ray telescope, XRT) data in their spectral analysis. 

\subsection{NuSTAR confirmation of BAT selected Compton-thick AGN}

It has been soon realised that the limited energy resolution of BAT may compromise the secure classification of Compton-thick AGN. In particular, it could be unclear whether a sources was a bonafide Compton-thick AGN or just a heavily obscured source with a column density bordering the Compton-thick regime. \cite{Tanimoto2022} systematically analysed the \nustar observations of 52 candidate Compton-thick AGN in the sample of \cite{Ricci2017}. They found that 28 objects are Compton-thick AGN at the 90\% confidence interval. Their analysis was based on the {\sc XCLUMPY} model of \cite{Tanimoto2019}. The Clemson University group has performed another search for Compton-thick AGN in the BAT catalogues \citep[e.g.][]{Marchesi2017,Marchesi2018, Marchesi2019, Torres-Alba2021, Zhao2020,Zhao2021,Silver2022,Sengupta2023}. Their candidate Compton-thick AGN were originally selected either because their properties (either their Seyfert 2 optical type or the absence of a bright ROSAT counterpart) suggest the presence of substantial intrinsic absorption or because they were lacking soft 0.3–7 keV coverage. The authors above used primarily the {\sc BORUS02} model of \cite{Balokovich2019} but also the {\sc MYTORUS} model of \cite{Murphy2009}. Despite the use of different models, the derived line-of-sight column densities agree reasonably well, with \cite{Tanimoto2022} within the uncertainties. \cite{Georgantopoulos2019} analysed the available \nustar observations of 19 sources from the \cite{Akylas2016} sample. The vast majority of these sources are common with the \cite{Ricci2015} sample. \cite{Georgantopoulos2019} use the absorption model of \cite{Murphy2009} together with a physically motivated thermal comptonisation model \citep{Titarchuk1994} to represent the coronal emission. They find that all but two sources are associated with Compton-thick obscuration. Additional \nustar observations of the six Compton-thick candidates of \cite{Akylas2016} which are not common with \cite{Ricci2015} show that three sources are indeed Compton-thick: 2MASX J00253292+6821442, ESO426-G002 and NGC4941. The other three (NGC 3081, NGC 3588NED01, ESO234-IG063) present lower obscuring column densities \citep{Traina2021, Silver2022}.

  \subsection{Models for Compton-thick absorption}
  We present below a very concise overview of the models most often used to quantify the absorption and reflection in the \nustar observations of BAT selected Compton-thick AGN. In Compton-thick objects, the obscuration is primarily because of scattering of X-ray photons on electrons and subsequent photoelectric absorption. There is a number of spectral codes available that utilise Monte Carlo simulations to model the absorption and reflection on the obscuring screen. 

The {\sc MYTORUS} \cite{Murphy2009} model assumes a tube-like azimuthally symmetric torus. The half-opening of the torus is assumed to be $60^\circ $ corresponding to  a scenario where there are equal numbers of obscured and unobscured AGN. The model can decouple the line-of-sight and equatorial column densities. This is important for taking into account continuum variability and time-delays between the direct (zeroth-order, or unscattered) continuum and the Compton-scattered continuum.

  However, it has become apparent that the AGN torus is composed of many individual clouds. This is primarily because the 10$\mu m$ silicate feature appears as both an emission and absorption feature in Seyfert-2 galaxies while the smooth torus models predict this feature only in absorption. In the infrared band \cite{Nenkova2008} constructed spectral models for clumpy tori. They assume a power-law in the radial direction and a normal distribution in the elevation direction for the configuration of the clouds. Following this model \cite{Tanimoto2019} constructed a new X-ray clumpy torus model designated as {\sc XCLUMPY} by adopting the same geometry of clump distribution. The clumpy torus models predict a higher fraction of unabsorbed reflection components as observed in many obscured AGNs. The line-of-sight column densities do not appear to differ significantly from the smooth tori model \citep[see discussion in][]{Torres-Alba2021}.

 One of the most widely used models is {\sc borus02} \citep{Balokovich2019}. The reprocessing medium is assumed to be a sphere of uniform density with conical cutouts at both poles, approximating a torus with variable covering factor. The half-opening angle of the polar cutouts, $\theta_{tor}$ is measured from the symmetry axis toward the equator, ranges from zero (full covering) to 83$^\circ$ corresponding to disk-like 10\% covering.  The line-of-sight component, $\rm N_H^{los}$, can have a different column density than the column density of the torus ($N_H^{tor}$), in order to account for clouds passing in front  of the line-of-sight. In practice, this feature  mimics the properties of a patchy torus.
  
Finally, \cite{Buchner2021} developed a clumpy torus named {\sc UXCLUMPY}. Naturally, the patchy geometry results in strong Compton scattering, causing soft photons to escape also along Compton-thick sight lines. This model introduces an additional Compton-thick reflector near the corona, a necessary feature in order to achieve acceptable spectral fits to the \nustar spectra. This additional component can be interpreted as part of the dust-free broad-line region, an inner wall or rim, or a warped disk.

\begin{table*}[h!]
\centering
\caption{The Compton-thick AGN sample}
\begin{tabular}{lcccccccc}
Name  & $\alpha$ & $\delta$ & redshift & $F_\mathrm{14-195\,keV}$ & $N_\mathrm{H}$ & $\log L_\mathrm{10-50\,keV}$ & Ref. & Sample \\
(1) & (2) & (3) & (4) & (5) & (6) & (7) & (8) & (9) \\
\hline
2MASX J00253292+6821442 & 00:25:33 & +68:21:44 & 0.012 & $18_{-4}^{+5}$ & $3.20_{-0.30}^{+0.35}$ & $43.06_{-0.12}^{+0.11}$ & g,h & 3 \\
MCG -07-03-007 & 01:05:27 & -42:12:58 & 0.02988 & $12\pm4$ & $0.90_{-0.09}^{+0.11}$ & $43.39_{-0.17}^{+0.13}$ & f,a & 1 \\
2MASX J01073963-113912 & 01:07:40 & -11:39:12 & 0.04746 & $10_{-10}^{+-10}$ & $3.00_{-0.05}^{+100.00}$ & $43.78_{-0.56}^{+0.99}$ & e & 2 \\
NGC 0424 & 01:11:28 & -38:05:01 & 0.01176 & $21\pm4$ & $3.20\pm0.19$ & $43.09_{-0.09}^{+0.08}$ & c,e,a,b & 1 \\
NGC 612 & 01:33:58 & -36:29:36 & 0.0298 & $55_{-4}^{+5}$ & $0.90\pm0.10$ & $44.05\pm0.04$ & e & 4 \\
NGC 1068 & 02:42:41 & -00:00:48 & 0.0038 & $35\pm4$ & $10.00_{-0.80}^{+100.00}$ & $43.06_{-0.27}^{+0.37}$ & c,r,b,l & 1 \\
NGC 1106 & 02:50:41 & +41:40:17 & 0.01447 & $19\pm5$ & $4.80_{-0.20}^{+100.00}$ & $43.60_{-0.27}^{+0.75}$ & a,p & 1 \\
2MFGC 02280 & 02:50:43 & +54:42:18 & 0.01515 & $27\pm5$ & $2.00_{-0.20}^{+2.17}$ & $43.34_{-0.11}^{+0.19}$ & a,b,h & 1 \\
NGC 1125 & 02:51:40 & -16:39:04 & 0.01093 & $17\pm4$ & $1.30_{-0.30}^{+0.70}$ & $42.74_{-0.13}^{+0.12}$ & c,a,e & 1 \\
NGC 1142 & 02:55:12 & -00:11:01 & 0.02885 & $88\pm5$ & $1.60\pm0.30$ & $44.33\pm0.05$ & e,c & 4 \\
NGC 1194 & 03:03:49 & -01:06:13 & 0.0136 & $37\pm5$ & $1.00\pm0.10$ & $43.20_{-0.06}^{+0.05}$ & e,c,b,a,g & 1 \\
ESO201-4 & 03:50:24 & -50:18:35 & 0.0359 & $21\pm4$ & $2.20_{-0.60}^{+0.50}$ & $43.98_{-0.11}^{+0.10}$ & e & 4 \\
CGCG 420-015 & 04:53:26 & +04:03:42 & 0.02939 & $28\pm5$ & $1.60_{-1.30}^{+1.20}$ & $43.85_{-0.20}^{+0.17}$ & c,a & 1 \\
ESO 005-G004 & 06:05:42 & -86:37:55 & 0.0062 & $34_{-4}^{+5}$ & $1.10\pm0.20$ & $42.49_{-0.07}^{+0.06}$ & c,a (*) & 1 \\
Mrk 3 & 06:15:36 & +71:02:15 & 0.0135 & $141\pm5$ & $0.87_{-0.08}^{+0.13}$ & $43.75\pm0.02$ & e & 2 \\
ESO 426-G002 & 06:23:46 & -32:13:00 & 0.0224 & $24\pm4$ & $1.00\pm0.03$ & $43.46_{-0.08}^{+0.07}$ & f & 3 \\
2MASX J06561197-4919499 & 06:56:12 & -49:19:50 & 0.041 & $12\pm4$ & $1.40_{-0.09}^{+0.16}$ & $43.77_{-0.16}^{+0.13}$ & a,f & 2 \\
MCG +06-16-028 & 07:14:04 & +35:16:45 & 0.01569 & $17\pm5$ & $1.00\pm0.17$ & $42.98_{-0.15}^{+0.12}$ & c,a,d,h & 1 \\
NGC 2788A & 09:02:39 & -68:13:37 & 0.01335 & $20\pm4$ & $2.31_{-0.76}^{+0.52}$ & $43.10_{-0.14}^{+0.11}$ & a,p & 1 \\
ESO 565-G019 & 09:34:44 & -21:55:40 & 0.01629 & $21\pm6$ & $3.00_{-0.70}^{+0.90}$ & $43.36_{-0.16}^{+0.15}$ & e,a,k & 2 \\
MCG +10-14-025 & 09:35:52 & +61:21:11 & 0.03937 & $8_{-3}^{+4}$ & $1.50_{-0.09}^{+0.53}$ & $43.57_{-0.23}^{+0.15}$ & a,r & 2 \\
NGC 3079 & 10:01:58 & +55:40:47 & 0.00372 & $33\pm4$ & $2.50\pm0.23$ & $42.22_{-0.06}^{+0.05}$ & c,a,b & 1 \\
ESO 317-G041 & 10:31:23 & -42:03:38 & 0.01932 & $17\pm4$ & $0.85_{-0.12}^{+0.16}$ & $43.14_{-0.13}^{+0.10}$ & a & 1 \\
NGC 3281 & 10:31:52 & -34:51:13 & 0.01061 & $86\pm5$ & $2.00\pm0.30$ & $43.49_{-0.05}^{+0.04}$ & e,g & 4 \\
NGC 3393 & 10:48:23 & -25:09:43 & 0.01251 & $26_{-5}^{+6}$ & $1.90_{-0.40}^{+1.50}$ & $43.13_{-0.13}^{+0.16}$ & c,a & 1 \\
NGC 4180 & 12:13:03 & +07:02:20 & 0.00699 & $17\pm5$ & $1.50_{-0.24}^{+0.36}$ & $42.35_{-0.15}^{+0.12}$ & e,a,p & 1 \\
ESO 323-G032 & 12:53:20 & -41:38:08 & 0.016 & $15_{-5}^{+6}$ & $1.80_{-0.49}^{+1.46}$ & $43.10_{-0.21}^{+0.19}$ & f,a & 2 \\
NGC 4941 & 13:04:13 & -05:33:06 & 0.0037 & $20\pm5$ & $3.10\pm1.70$ & $42.06_{-0.21}^{+0.19}$ & b,j & 3 \\
NGC 4945 & 13:05:27 & -49:28:06 & 0.01932 & $285\pm6$ & $3.80\pm0.16$ & $44.74\pm0.02$ & c,a,o & 1 \\
Circinus & 14:13:10 & -65:20:21 & 0.0014 & $272\pm4$ & $7.00_{-2.00}^{+3.00}$ & $42.69_{-0.17}^{+0.20}$ & n,b & 1 \\
NGC 5643 & 14:32:41 & -44:10:28 & 0.004 & $18_{-5}^{+6}$ & $1.60_{-0.30}^{+0.40}$ & $41.90_{-0.16}^{+0.13}$ & c,a,b & 1 \\
NGC 5728 & 14:42:24 & -17:15:11 & 0.00935 & $89\pm6$ & $1.20_{-0.04}^{+0.11}$ & $43.28\pm0.03$ & a,b,c & 1 \\
CGCG 164-019 & 14:45:37 & +27:02:05 & 0.0299 & $14_{-4}^{+5}$ & $1.20_{-0.36}^{+0.50}$ & $43.51_{-0.17}^{+0.14}$ & c,a,h & 1 \\
ESO 137-G034 & 16:35:14 & -58:04:48 & 0.00914 & $28_{-5}^{+6}$ & $2.90\pm0.20$ & $42.97_{-0.10}^{+0.08}$ & a,b & 1 \\
NGC 6232 & 16:43:20 & +70:37:57 & 0.0148 & $12\pm4$ & $0.57_{-0.10}^{+0.70}$ & $42.72_{-0.18}^{+0.15}$ & e,a,b & 1 \\
NGC 6240 & 16:52:59 & +02:24:03 & 0.02448 & $72\pm6$ & $1.20\pm0.14$ & $44.04\pm0.04$ & e,a & 1 \\
NGC 6552 & 18:00:07 & +66:36:54 & 0.02649 & $19\pm4$ & $2.20_{-0.31}^{+0.40}$ & $43.68_{-0.11}^{+0.09}$ & f,a & 1 \\
NGC 6921 & 20:28:29 & +25:43:24 & 0.0145 & $78\pm5$ & $1.70\pm0.30$ & $43.68\pm0.05$ & m,h & 1 \\
ESO 464-G016 & 21:02:24 & -28:10:29 & 0.03635 & $17_{-5}^{+6}$ & $0.86_{-0.16}^{+0.17}$ & $43.72_{-0.16}^{+0.13}$ & e,a & 2 \\
NGC 7130 & 21:48:20 & -34:57:04 & 0.01615 & $16_{-5}^{+6}$ & $3.40_{-0.40}^{+0.50}$ & $43.30_{-0.19}^{+0.14}$ & e,a,b & 1 \\
NGC 7212 & 22:07:01 & +10:13:52 & 0.02667 & $11_{-4}^{+5}$ & $1.90_{-0.40}^{+0.50}$ & $43.42_{-0.22}^{+0.16}$ & c,a & 1 \\
NGC 7479 & 23:04:57 & +12:19:22 & 0.00794 & $20\pm5$ & $6.10\pm0.90$ & $43.00_{-0.14}^{+0.12}$ & c,a,d & 1 \\
CGCG 475-040 & 23:07:49 & +22:42:37 & 0.03473 & $14_{-4}^{+5}$ & $1.60_{-0.15}^{+0.23}$ & $43.71_{-0.17}^{+0.12}$ & f,a & 2 \\
NGC 7582 & 23:18:24 & -42:22:14 & 0.0053 & $81\pm4$ & $1.70_{-0.15}^{+0.10}$ & $42.81\pm0.03$ & c,a,e & 1 \\
 \hline
\end{tabular}
  \label{sample}
\tablefoot{Column Description: (1) Name (2) Right Ascension (3) Declination (4) Redshift (5) Observed flux (14-195~keV) from BAT catalogue in units of $10^{-12}\,\mathrm{ergs\,sec^{-1}\,cm^{-2}}$ (6) Hydrogen column density in units of $10^{24}\,\mathrm{cm^{-2}}$ (7) log Luminosity (10-50~keV) (8) References: a. \cite{Tanimoto2022} b. \cite{Georgantopoulos2019} c. \cite{Marchesi2018} d.\cite{Marchesi2019} e. \cite{Zhao2021} f. \cite{Torres-Alba2021} g. \cite{Semena2019} h. \cite{koss2016} i. \cite{Panagiotou2021} j. \cite{Jana2022} k. \cite{Traina2021} l. \cite{Zaino2020} m. \cite{Yamada2021} n. \cite{Arevalo2014} o. \cite{Puccetti2014} p. \cite{Sengupta2023} q. \cite{Marinucci2016} r. \cite{Oda2017} s. \cite{Zhao2020}  
 (9) Parent sample: 1. Common in \cite{Ricci2015} and \cite{Akylas2016};
  2. Only in \cite{Ricci2015}; 3. Only in \cite{Akylas2016} 4. \cite{Zhao2021}}
 \end{table*}

\section{The Sample}\label{thesample}
\subsection{Sample Selection}
We compile our sample of Compton-thick AGN in the local Universe based on the \swift BAT 70-month survey. We start by using the Compton-thick AGN samples selected by \cite{Ricci2015} and \cite{Akylas2016}. Finally, we complement our sample with a few additional Compton-thick sources reported by the Clemson group (see Table\ref{sample}). The condition for inclusion of these sources in our sample is that they are detected in the 70-month survey and thus they are included in the sample of \cite{baumgartner2013}. These evaded classification by both \cite{Ricci2015} and \cite{Akylas2016}. As \nustar observations exist for all these sources, there is an accurate determination of the column density available. We include all sources with column densities consistent with $\rm N_H=10^{24} cm^{-2}$ at the 90\% confidence level. Note that the selected threshold value is somewhat below the column density that corresponds to $\tau=1$ for Compton scattering. We include only sources with redshift $z<0.05$ corresponding roughly to 200 Mpc. Our  sample contains 45 sources and is given in Table \ref{sample}. The first reference in column (8) of Table \ref{sample} denotes the origin of the column density. All our sources have a column density much lower than $\rm 10^{25} ~cm^{-2}$ with the exception of 
 NGC 1068 which has a border-line column density of 
 $\rm N_H= 10^{+100}_{-0.80}\times 10^{25}~cm^{-2}$. We chose to exclude this source and restrict our analysis only to the $\rm 10^{24}-10^{25}~cm^{-2}$ sources. This is because most  current models for Compton-thick absorption 
  do not produce reliable results at these extreme column densities. 

The majority of the column densities comes from the work of either \cite{Zhao2021}, \cite{Marchesi2018} or alternatively \cite{Tanimoto2022}. In the vast majority of sources, the differences between the column densities are consistent within the uncertainties. An exception is  ESO 005-G004 where \cite{Tanimoto2022} using {\sc XCLUMPY} find that the source is heavily obscured while \cite{Marchesi2018} using {\sc MYTORUS} find that the source is marginally Compton-thick. \cite{Zhao2021} using {\sc BORUS02} find a column density of $\rm N_H\approx 5\times10^{24} ~cm^{-2}$. The differences are probably attributed to the fact that the different models prefer a significantly different value for the photon-index which counteracts the derived column density. Finally, we note that the luminosities are estimated in a consistent manner as described in the section below. 
  
\subsection{Estimation of absorption-corrected luminosities}

The selected sample of local Compton-thick AGN has been studied with a variety of spectral models in the literature.
While estimations of the line of sight column density, $\rm N_H$ are reasonably robust and model independent \cite[e.g][]{Saha2022} if X-ray data above 10 keV is included, the remaining parameters can be quite uncertain.
This can cause significant discrepancies particularly when estimating the absorption corrected luminosity.
For consistency we have assumed a single X-ray spectral model for all the sources. Given this model, the observed BAT flux, and $\rm N_H$ estimates from the literature, we calculated the corresponding intrinsic luminosity for each AGN in our sample. We used the spectral model assumed by \cite{Ananna2022} in their study of the X-ray luminosity function of \nustar sources. It is a modification of the model assumed by \cite{Ueda2014}, where the torus absorption and reflection is now modelled using {\sc BORUS02}. For Compton-thick sources, a photon index of $\Gamma=1.8$ is used with an energy cut-off of 200~keV, an inclination angle of $72 \deg$, and a half-opening angle of $60 \deg$. In order to check the validity of our derived luminosities, we also used the {\sc XCLUMPY} model. We find no significant differences in our derived luminosities.

For a rigorous treatment of the $\rm N_H$ and flux uncertainties, for each source we generate 2000 random samples of $\rm N_H$, flux pairs and calculate the corresponding luminosity. The sampling is done assuming a generalized extreme distribution parameterized in such a way that the its median corresponds to the quoted value of $\rm N_H$/flux in Table \ref{sample}, and the 5 and 95 percentiles correspond to the respective error interval. This distribution is recommended to reproduce strong asymmetric confidence intervals \cite{Possolo2019}.

\subsection{Detection probability and selection biases}\label{area}

The sensitivity curve (flux limit vs. area covered) of the BAT survey is not uniform across the sky \citep{baumgartner2013}. Towards the flux limit bright end (low sensitivity), the area covered is the largest but a considerable number of sources are less likely to be detected. This is because they will present faint flux because 
they will have low luminosity or higher column density or simply they are found at higher redshift. This introduces some biases that have to be addressed and quantified in order to estimate accurately the X-ray luminosity function of the intrinsic population of the Compton-Thick sources. Then we need to estimate the probability that a Compton-thick source with column density, $\rm N_H$, redshift $z$ and intrinsic luminosity $\rm L_X$ will be detected in the BAT survey. To this end, we used the {\sc BORUS02} model as above to build the sensitivity maps. We assumed a photon index of $\Gamma=1.8$, inclination angle $i=72\deg$  and torus half-opening  angle $\sigma=60\deg$. Then, we calculated the expected flux at a given set of $z$, $\rm L_X$ and $\rm N_H$ using a grid of 50 bins in each parameter. The probability of detecting a source was derived by convolving the expected flux with the area curve of BAT \citep{baumgartner2013}. The upper and lower panels of Fig. \ref{senmap} present the detection probability maps as a function of redshift and intrinsic X-ray luminosity for a source with  $\rm N_H\sim10^{24}~cm^{-2}$ and $\rm N_H\sim10^{25}~cm^{-2}$, respectively.

 \begin{table}
\caption{Best-fit values and $2\sigma$ errors for the parameters of the luminosity function.}   \label{XLFparameters}      
\centering                                      
\begin{tabular}{l  c  c }          
\hline\hline                        

Parameter  & Prior & Best value \\

\hline                                   
\rule{0pt}{2ex} 
        $\log A$         & -7, -3 & $-4.40^{+0.40}_{-0.50}$ \\ \rule{0pt}{3ex}  
        $\log L_*$       & 42, 46 & $42.86^{+0.44}_{-0.43}$  \\ \rule{0pt}{3ex}
        $\gamma_1$      & -2,  2&   $0.01^{+0.51}_{-0.74}$     \\ \rule{0pt}{3ex}
        $\gamma_2$      & 1,  6&    $1.72^{+0.40}_{-0.37}$     \\ 
        \hline                                 
\end{tabular}
\tablefoot{The normalisation $A$ and the break luminosity $\rm L_*$ are given in units of $\rm Mpc^{-3}$ and $\rm erg~s^{-1}$, respectively.}
\end{table}

\begin{table}
\caption{Compton-thick AGN fractions}   \label{fractions}      
\centering                                      
\begin{tabular}{l  c  c } 
\hline 
Sample  & fraction & redshift\\
\hline
   $^{\cross}$ &      0.24$^{+0.05}_{-0.05}$  & $<$0.05 \\
\cite{Burlon2011} $^{\cross}$ & $0.20^{+0.09}_{-0.06}$ & $<$0.1 \\
\cite{Ricci2017} $^{\cross}$ & $0.27^{+0.04}_{-0.04}$ & $<$0.1 \\
\cite{Torres-Alba2021}$^{\cross}$ & $0.20^{+0.05}_{-0.05}$ & $<$0.01 \\
\cite{Ueda2014}  $^{\cross}$ &      $0.27^{+0.05}_{-0.05}$  & $<$0.1  \\
\cite{Buchner2015} $^{\dagger}$ &  $0.43^{+0.10}_{-0.10}$ & $\approx0$ \\
\cite{Laloux2023} $^{\dagger,1}$ &  $0.21^{+0.16}_{-0.10}$ &  $<$0.5 \\
\cite{Akylas2024} $^{2}$ &  $0.25^{+0.05}_{-0.05}$ &    z$<$0.02 \\
\cite{Boorman2024} $^{3}$ & $0.35^{+0.06}_{-0.06}$ & $<$0.044 \\
\cite{Ananna2019}  $^{4}$  &  $0.33^{+0.10}_{-0.10}$ & $<$0.1 \\
\hline                          
\end{tabular}
\tablefoot{Column Description: (1) Sample (2) Compton-thick fraction ($\rm N_H=10^{24-25} cm^{-2}$)
(3) redshift; $\cross$: fractions based on BAT in the local Universe; $\dagger$ fractions based on \xmm and \chandra at higher redshifts;
 $^1$ based on mid-IR priors $^2$ AGN sample selected using {\it WISE}  warm colours $^3$ AGN sample selected using warm {\it IRAS} colours $^4$ based on X-ray background synthesis models} 
\end{table}

\subsection{Column Density, redshift, luminosity distributions}
  The redshift distribution of our sample is given in Fig. \ref{distributions}. The redshift distribution peaks at 
  $z\approx 0.015$. In the same figure we present the distribution of the unabsorbed luminosity in the 10-50 keV band as well as the hydrogen column density, $\rm N_H$, distribution. The column density distribution is 
  dominated by sources just above the threshold of $\rm N_H=10^{24} cm^{-2}$ while there is a significant deficit of sources with $\rm \log(N_H)[cm^{-2}]>24.5$. This can be most probably attributed to a selection effect. The widest possible range of column densities can be detected only at the lowest redshifts owing to the limited sensitivity of the BAT survey. This is evident in Fig. \ref{z-lx} where we plot the $\rm N_H$ values vs the redshift of each source. It is apparent that there is a strong correlation between $\rm N_H$ and redshift in the sense at the highest redshifts we can detect only Compton-thick AGN  with relatively low column densities. In the previous section we presented the detection probability as a function of intrinsic X-ray luminosity and redshift for a given column density. From Fig. \ref{senmap} we see that sources like NGC1068 with intrinsic luminosity $\rm L_X \approx10^{44} erg~s^{-1}$ and column density $\rm N_H\sim10^{25}cm^{-2}$ can be detected (detection probability 
    $\rm P_{detection}>0.5$) up to a redshift of $z=0.03$. A heavily obscured Compton-thick source with $\rm N_H=10^{25} cm^{-2}$ must have an intrinsic luminosity of $\rm L_X > 3\times 10^{44} erg~s^{-1}$ in order to be detected at the redshift limit of our survey $z=0.05$.

\section{Analysis}\label{analysis}

\subsection{X-ray luminosity function}\label{xlf}

We define the differential X-ray luminosity function, $\phi$, as the number of sources $N$ per comoving volume $V$ and per logarithmic interval $\rm logL_{X}$ as a function of redshift, $z$, and luminosity, $\rm L_X$

\begin{equation}
\phi(L_X,z) = \dv{\Phi(L_X,z)}{logL_X} = \dfrac{\dd[2]{N}(L_X,z)}{\dd V \dd logL_X}.
\end{equation}

Following \cite{Akylas2016} and \cite{Ananna2022}, we model 
 the  differential luminosity function  with a broken power-law \citep{Maccacaro1984,barger2005}, defined as:
\begin{equation}
\dv{\Phi(L_X)}{logL_X} =
A \times \left[\left(\dfrac{L_X}{L_*}\right)^{\gamma_1} + \left(\dfrac{L_X}{L_*}\right)^{\gamma_2}\right]^{-1},
\end{equation}
where $A$ is a normalization factor, $L_*$ is the break luminosity, while $\gamma_1$ and $\gamma_2$ are the slopes of the power-law at the faint-end and the bright-end slope respectively. 

Furthermore, we calculated the binned luminosity function for visualisation purposes following the \citet{Page2000} method that is based on the $\rm 1/V_{max}$ method \citep{Schmidt1968,Avni1980}. After dividing the sample of Compton-thick sources into redshift, luminosity and hydrogen column density bins, the binned luminosity function can be estimated such as:
\begin{equation}
\phi(L_X,z,N_H) = \dfrac{\langle N \rangle}{\iiint\Omega(L_X,z,N_H) \dv{V}{z}\, \dd log L_X\,\dd log N_H~\dd z} ,
\end{equation}
where $\langle N \rangle$ is the number of sources in each bin, $\dv*{V}{z}$ is the differential comoving volume, and $\Omega$ represents the survey area (Sect.~\ref{area}).

\subsection{Fit and parameter estimation}\label{fit}
 We use Bayesian inference to estimate the parametric form of the X-ray luminosity function, following the approach of 
 \cite{Loredo2004}.
 The full description of the method is given in \cite{Pouliasis2024}. Here, we just outline 
 the basic concepts.
Given a data-set of $n$ observations, $D = \{d_i; i=1,...,n\}$, and a model for the X-ray luminosity function defined by a set of parameters $\vb*{\Theta}$, according to the Bayes' theorem
 the posterior probability i.e. the probability of obtaining the 
 selected model given the observational data is 
\begin{equation}
P(\vb*{\Theta} | D) = \dfrac{P(D|\vb*{\Theta}) P(\vb*{\Theta})}{P(D)}.
\end{equation}

 The likelihood, $\La = P(D | \vb*{\Theta})$,  is the probability of obtaining the observational data given the model; $P(\vb*{\Theta})$, the prior, is the {\it a priori} probability for the parameters of the model. $P(D)$ is the evidence of the model: $P(D)= \int P(\vb*{\Theta}| D )d\vb{\Theta}$.
We derive the posterior probability distribution of the model parameters, using the nested-sampling Monte Carlo algorithm MLFriends \citep{Buchner2017}, implemented in the \texttt{UltraNest} package. Nested sampling algorithms allow tracing the posterior distribution of the model, given a data set, while at the same time calculating the Bayesian evidence.
 The Bayesian approach allows for a rigorous treatment of the uncertainties in the X-ray properties  of the sources.  During the inference process, we assumed flat priors for the model parameters, either uniform or log-uniform, that span a reasonably broad range of the parameter space according to previous studies in the literature \citep{Akylas2016,Ananna2022}. The range of our priors 
 is consistent with the $1/V_{max}$ non-parametric luminosity function. In Table~\ref{XLFparameters} we provide the minimum and maximum values allowed in the flat priors we used for the parameters of our X-ray luminosity function model.

The log-likelihood of this process can be written as:
\begin{multline}\label{eq:xlflikelihood}
\ln \La(\{d_i\} | \vb*{\Theta}) = \\
\shoveleft-\lambda + \sum_i \ln\iiint P_i(\LX, z, \mnh | \vb*{\Theta})~\dv{V}{z}\dlog\mnh~\dlog\LX~\dd z.
\end{multline}

The parameter $\lambda$ is the expected number of observed sources for a Poisson process, given an XLF model with parameters $\vb*{\Theta}$.
\begin{equation}\label{eq:expectedsources}
\lambda = \iiint \phi (\LX, z, \mnh | \vb*{\Theta}) \Omega(\LX, z, \mnh)~\dv{V}{z}\dlog\mnh~\dlog\LX~\dd z,
\end{equation}
where $\Omega$ is the survey sensitivity function and $\phi$ is the luminosity function.

The parameter $P_i$ in Eq.~\ref{eq:xlflikelihood} is given by:
\begin{equation}
P_i(\LX, z, \mnh | \vb*{\Theta}) = p(d_i | \LX, z, \mnh)~\phi(\LX, z, \mnh | \vb*{\Theta})~\Omega(\LX, z, \mnh).
\end{equation}
where $p(d_i | \LX, z, \mnh)$ is the probability of the source $i$ being at redshift $z$ with column density $N_H$ and luminosity $L_X$. This probability is given by the posterior probability distributions we obtained during the X-ray spectral analysis. We have included in this term the sensitivity function of the survey $\Omega$.
The integral involving $P_i$ can be calculated using an importance sampling  Monter Carlo integration technique \citep{Press1992}.
The integration limits used in Eqs.~(\ref{eq:xlflikelihood})  are [0,0.05], [41.0,46.0] and [23.5,26.0] for the parameters $z$, $\log L_X$ and $\log N_H$, respectively.

\begin{figure*}
\center
    \includegraphics[width=1\textwidth]{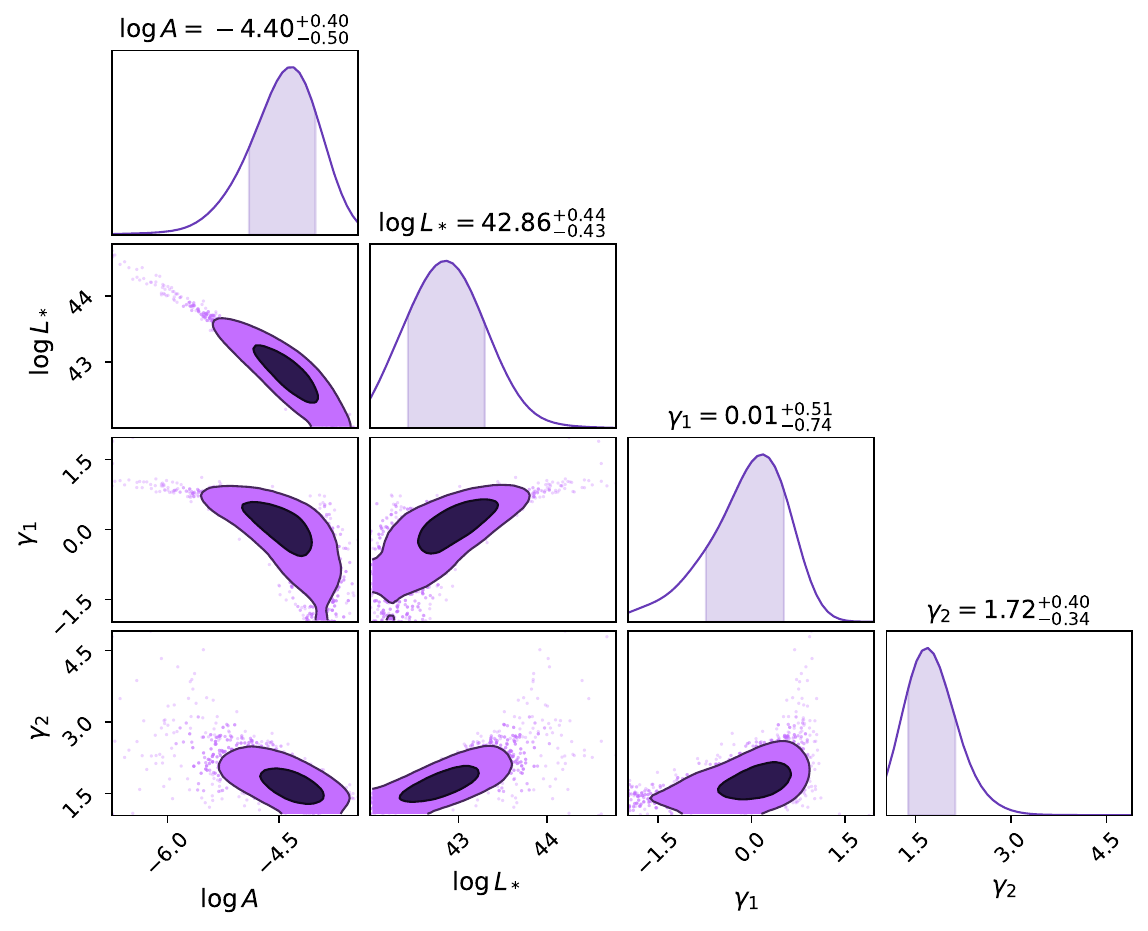} 
\caption{One-dimensional (diagonal panels) and two-dimensional marginalised posterior distributions for the double-power law model parameters. The shaded areas in the 2D posterior distributions correspond to $1\sigma$ and $2\sigma$ confidence levels (2D values, i.e. 39\% and 86\% respectively). The shaded areas for the 1D posteriors correspond to $1\sigma$ confidence level.}\label{diagonal}
\end{figure*}

\begin{figure*}

\begin{tabular}{cc} 

\includegraphics[width=0.45\textwidth]{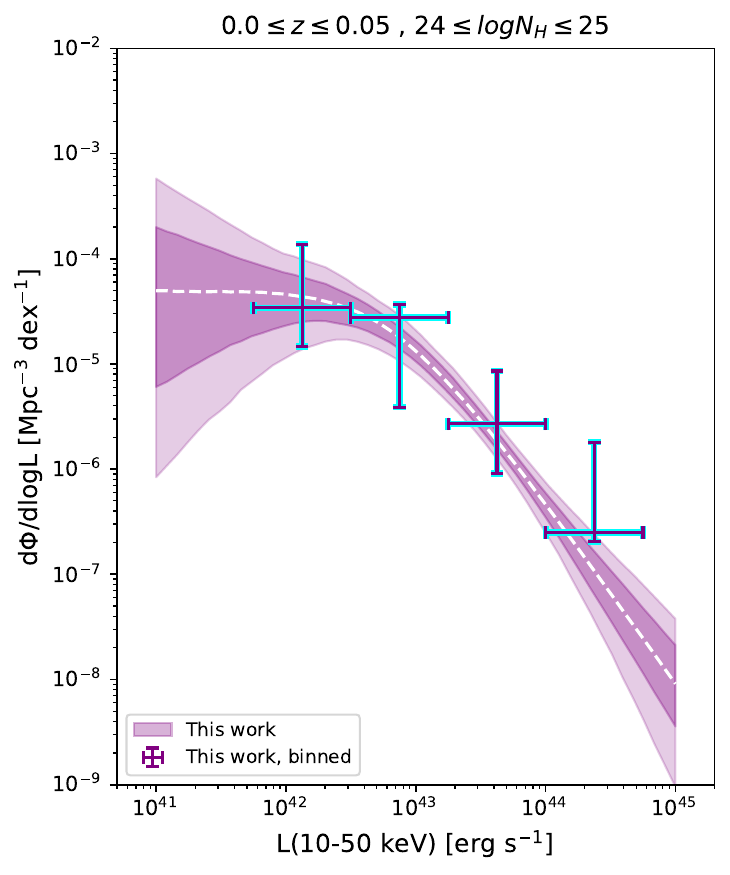}  &
\includegraphics[width=0.45\textwidth]{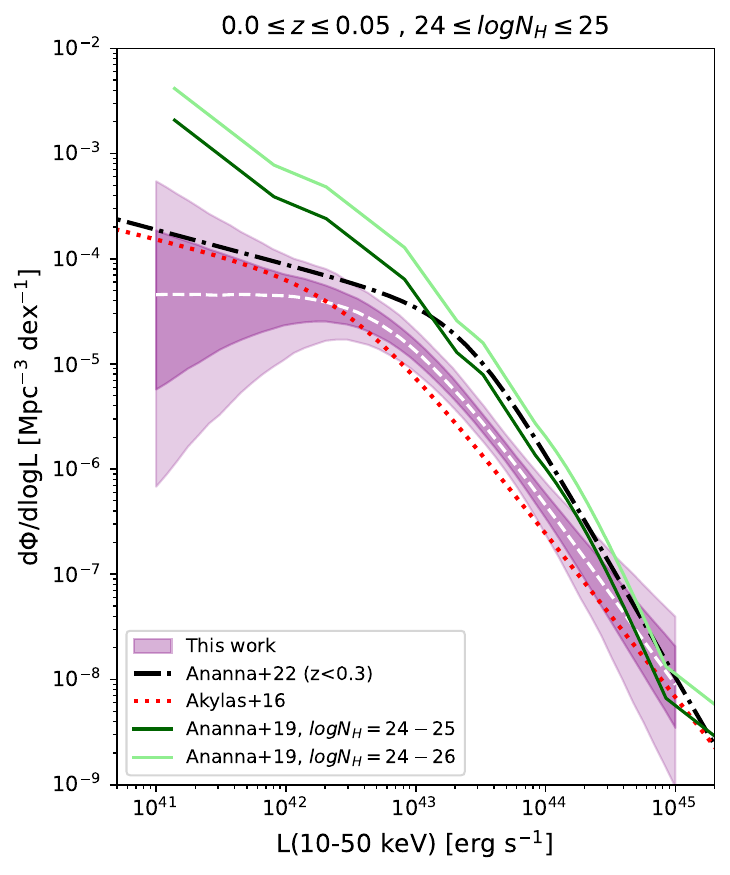}  

\end{tabular}
\caption{Left panel: The Compton-thick ($\rm logN_H[cm^{-2}]=24-25$) X-ray luminosity function in the redshift range $0.0\leq z \leq 0.05$. The shaded regions represent the 68\% and 95\% confidence intervals. The points show the binned $\rm 1/V_{max}$ luminosity function with the corresponding  68\% uncertainties. Right panel: Our luminosity function compared with those of \citep{Akylas2016,Ananna2022} and also by the population synthesis models of \citet{Ananna2019}. The best-fitting model of the latter is evaluated at the mean redshift of our analysis ($z=0.025$).}\label{plot_XLF_zbins}
\end{figure*}

\begin{figure}
\centering
\includegraphics[width=0.47\textwidth]{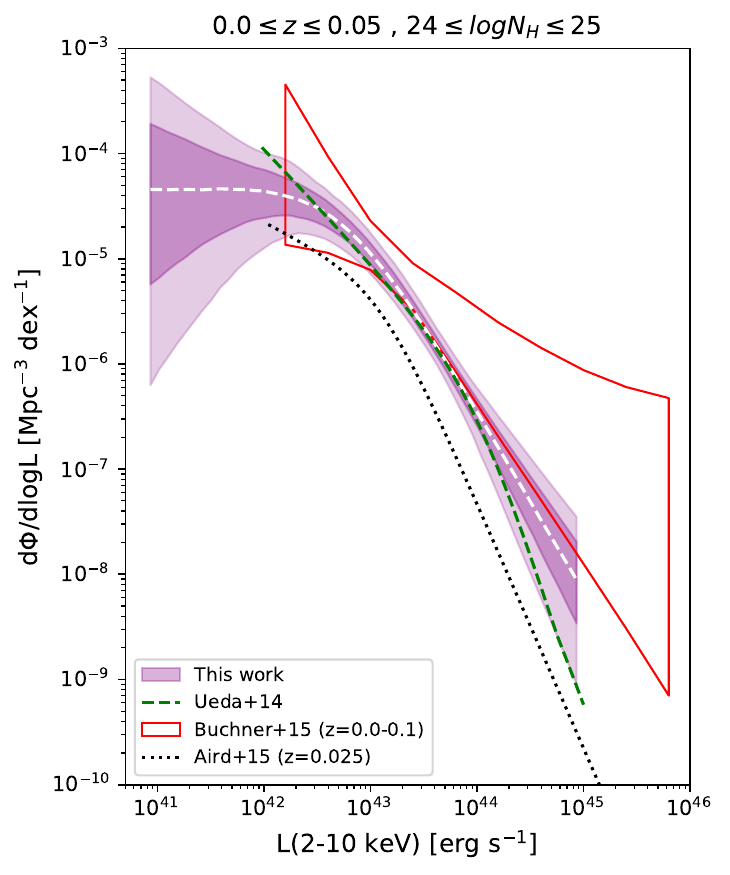}  
\caption{The X-ray luminosity function ($\rm logN_H[cm^{-2}]=24-25$) in the 2-10 keV band. The shaded regions represent the 68\% and 95\% confidence intervals. The green dash line corresponds to the Compton-thick luminosity function of 
\cite{Ueda2014}. 
 The dotted line denotes the luminosity function of 
\cite{Aird2015a}. Finally, the red fish-shape diagram corresponds to the luminosity function of \cite{Buchner2014} (see text for details).}
\label{2_10}
\end{figure}

 \begin{figure}
\centering
\includegraphics[width=0.50\textwidth]{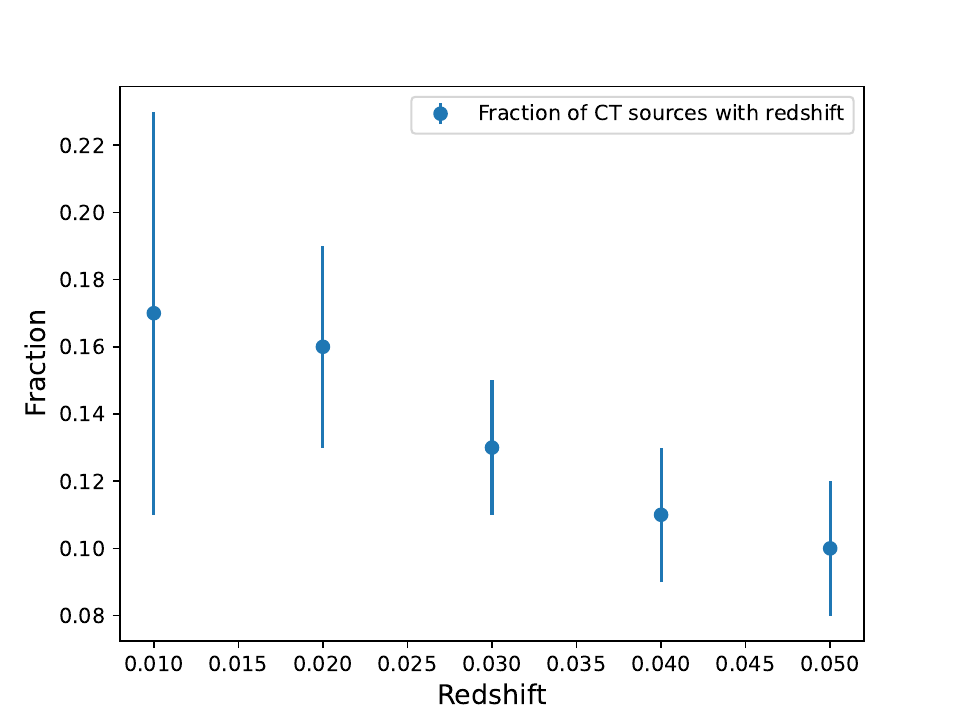}\\
\caption{The {\it observed} fraction of Compton-thick AGN as a function of redshift.}
\label{fraction}
\end{figure}

\begin{figure*}
\centering
\includegraphics[width=0.49\textwidth]{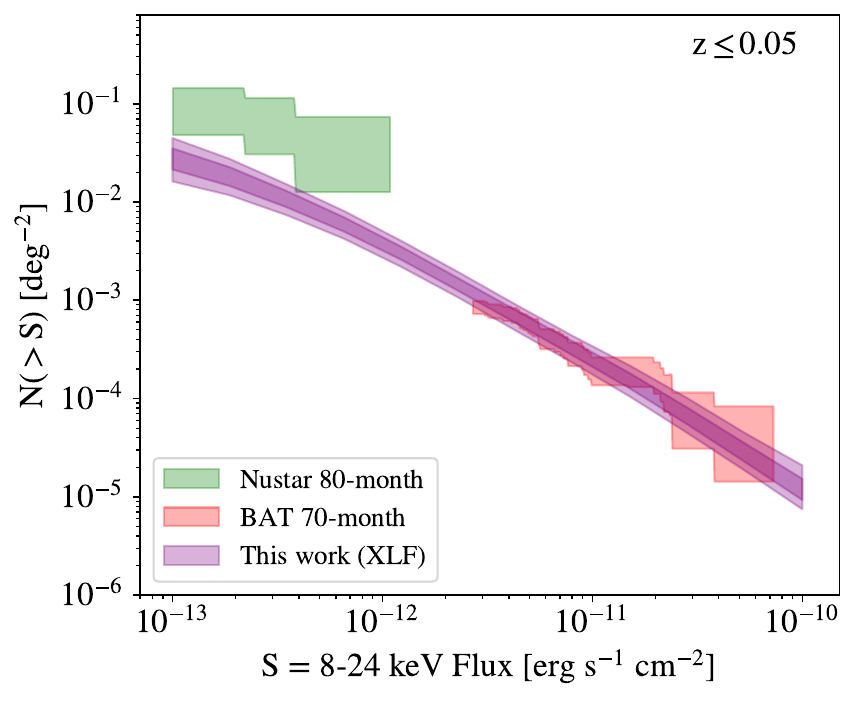}  
\includegraphics[width=0.49\textwidth]{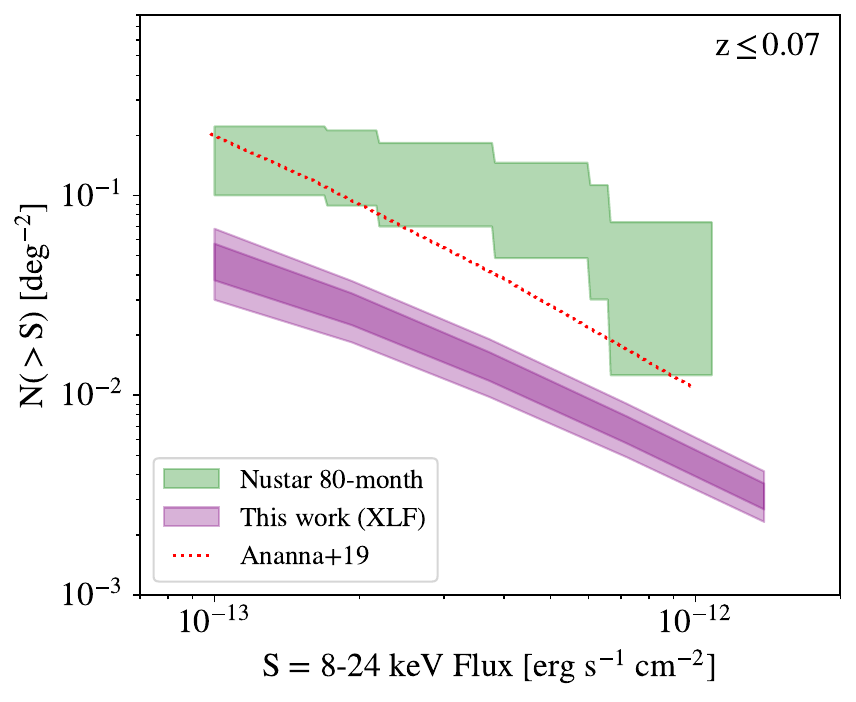}
\caption{The observed and predicted number counts in the  8-24 keV band. a) Left: redshift range $z<0.05$. The BAT 70-month survey 
 number counts are plotted (red shaded diagram) together with the 
 predictions of our luminosity function (shaded  curve). The \nustar number counts
 are shown with the green diagram b) Right: redshift range $z<0.07$.
  The green shaded diagram corresponds to the \nustar number counts.
   The predictions of our luminosity function are depicted with the shaded curve. The red dotted line denotes  the predictions 
   of the X-ray background synthesis model of \cite{Ananna2019} in the 
   $\rm 10^{24}-10^{26}~cm^{-2}$ column density range.}
\label{NumberCounts}
\end{figure*}

\section{Results}

\subsection{The 10-50 keV X-ray Luminosity function}
Here, we present the best-fit results and the confidence intervals for the parameters of the luminosity function (Eq. 2). These are the normalization, A, the break $L_*$, and the faint $\gamma_1$ and bright-end slopes $\gamma_2$. The values together with the 68\% confidence intervals are given in Table \ref{XLFparameters}. In Fig. \ref{diagonal}, we present the corner plots. These depict the relationships between various parameters by drawing a scatter plot of every set of parameters e.g. $L_\star$ vs. $\gamma_1$. The diagonal boxes illustrate the marginalised posterior distribution of each parameter. The break luminosity is $\rm logL_\star[erg~s^{-1}]\approx43$ while the faint-end of the luminosity function is quite flat having a slope of $\gamma_1=0.01^{+0.51}_{-0.74}$. This suggests that may be no ample room for a faint population of Compton-thick AGN. 
  
In Fig. \ref{plot_XLF_zbins}, we present the X-ray luminosity function in the 10-50 keV band. In the left panel we compare the parametric luminosity function with the non-parametric one derived using the $\rm 1/V_{max}$ method. In the right panel of Fig. \ref{plot_XLF_zbins}, we compare our luminosity function with those of \cite{Akylas2016} and \cite{Ananna2022}. These luminosity functions have been derived in the 20-40 keV and 14-195 keV band, respectively. We homogenised the different luminosity functions to the 10-50 keV band using the corresponding spectral models assumed by each work.

The luminosity function of \cite{Ananna2022} lies somewhat above our estimate especially around $\rm L_\star$. It has to be noted that the two luminosity functions contain a different set of objects despite the fact that they both come from the 70-month BAT survey. Moreover, it has to be taken into account that the luminosity function of \cite{Akylas2016} is not corrected for obscuration. In Fig. \ref{plot_XLF_zbins} (right panel), we also compare with the X-ray background synthesis models of \cite{Ananna2022}. These refer to both the Compton-thick population with column densities $\rm N_H=10^{24-25}~cm^{-2}$ and $\rm N_H=10^{25-26}~cm^{-2}$, respectively. We can compare our luminosity function only with the former as they refer to the same column density range. It appears the model predicts a very high number of Compton-thick AGN at low luminosities, a few times $\rm 10^{42}~erg~s^{-1}$. This is rather at odds with our findings and even with the luminosity function of \cite{Ananna2022}.

\subsection{Comparison with the 2-10 keV luminosity function} \label{lf_2_10}
  Next, we compare our luminosity function with other Compton-thick luminosity functions derived in the softer 2-10 keV band. \cite{Ueda2014} constructed the luminosity function in the  2-10 keV band compiling data from various missions operating in these wavelengths. They also used the BAT 9-month survey which contains a number of known Compton-thick AGN to determine their fraction in the local Universe. Their luminosity function has been used to model the spectrum of the X-ray background in the energy range 1-1000 keV resulting in a very good agreement with the data.   
 In Fig. \ref{2_10}, we compare the \cite{Ueda2014} with our luminosity function in the 2-10 keV band. We plot here the luminosity function of \cite{Ueda2014} that refers to the $\rm 10^{24-25}~cm^{-2}$ column density range.
  For the conversion of our luminosity function  in the 2-10 keV band we use the same spectral model used in sections 3.2 and 3.3. 
Their luminosity function is in good agreement with ours at luminosities brighter than $\rm L_X\approx10^{42}~erg~s^{-1}$. However, the former appears to present an upturn at faint luminosities.  

In Fig. \ref{2_10} we also plot the luminosity function 
 derived by \cite{Buchner2015} as a fish-shape diagram.
 This region gives the 10\%-90\% quantiles as a range for the uncertainty of the luminosity function.
 Note that the luminosity function of \cite{Buchner2015} typically refers to the $\rm N_H=10^{24-26}~cm^{-2}$ column density range. 
  However, the reflection-dominated sources with 
  $\rm N_H=10^{25-26}~cm^{-2}$
  are very sparsely sampled because of the extreme
  obscuration \citep[e.g.][]{Buchner2015}. It is therefore reasonable to assume that the luminosity
  functions of \cite{Buchner2015} quoted in the 
   $\rm N_H=10^{24-26}~cm^{-2}$ column range provide very good 
  approximations to the luminosity function in the 
  $\rm N_H=10^{24-25}~cm^{-2}$  column density range.
  Finally, we plot the luminosity function derived by \cite{aird2015b} for the $\rm N_H=10^{24-25}~cm^{-2}$ column density range. 

 \subsection{The observed fraction of Compton-thick AGN}
 We estimate the observed cumulative fraction of Compton-thick AGN in different redshift bins. For a given redshift, $z_i$, this is defined as the ratio of the Compton-thick AGN up to $z_i$ over the total number of (non-blazar) AGN in the same redshift bin. The fraction is given in Fig. \ref{fraction}. The fraction is the highest in the first redshift bin $z<0.01$ with $f\approx0.17\pm0.06$. According to Fig. \ref{senmap} an obscured source with $\rm N_H\approx10^{24} cm^{-2}$ and $\rm L_X < 3\times10^{42} erg~s^{-1}$ cannot  be easily detected even in our first redshift bin i.e. up to $\rm z=0.01$ ($\rm \approx 43 Mpc$). This is exacerbated at higher column densities. An AGN that is obscured by a column density of $\rm 10^{25}~cm^{-2}$ can be detected at z=0.01 
 only if it has a luminosity above $\rm L_X>10^{43}~erg~s^{-1}$. This means that our estimated fraction $f=0.17$ can be considered only as a lower limit to the fraction of Compton-thick AGN.

\subsection{The intrinsic fraction of Compton-thick AGN based on the luminosity function}
Our obscuration corrected luminosity function   
provides us with the opportunity to estimate the intrinsic fraction 
of Compton-thick AGN. Towards this end, we estimate the {\it intrinsic} number of Compton-thick AGN in the 2-10 keV band using the following expression.
\begin{equation}
\rm \int_{z=0}^{z=0.05} \int_{logL_X=42}^{logL_X=45} \Phi(L) 
~\dv{V}{z}~\dlog\LX~\dd z
\end{equation}

Our luminosity function 
predicts 112 Compton-thick with luminosity $\rm L_X>10^{42} ~erg~s^{-1}$ (10-50 keV) over the whole sky up to  redshift of $z=0.05$. 
Then we compare with the 
number of Compton-thin AGN ($\rm N_H=10^{20-24}~cm^{-2}$) derived 
using the above expression and the 
\cite{Ueda2014} Compton-thin luminosity function. 
The number of Compton-thin AGN is 359.
The fraction of Compton-thick sources is defined as:
\begin{equation}
f_{24-25} = N_{24-25}/(N_{20-24}+N_{24-25})
\end{equation}
where $\rm N_{24-25}$ and $\rm N_{20-24}$ are the
numbers of objects in the column density range 
$\rm N_H=10^{24-25} cm^{-2}$ and 
 $\rm N_H=10^{20-24} cm^{-2}$ and they are derived 
 from the current work and \cite{Ueda2014} respectively. 
Then the fraction of Compton-thick AGN vs. the total number of AGN is 24$\pm$5\%. The error is derived by sampling the uncertainty space 
of our Compton-thick luminosity function parameters
at the 68\% confidence level.

Next, we compare our findings with the Compton-thick fractions derived in the literature. 
 \cite{Ricci2015} have derived a fraction of 27$\pm$4\% by modelling the absorption distribution of the $\rm N_H=10^{24-25}~cm^{-2}$ Compton-thick AGN from the BAT 70-month survey. This figure is entirely compatible with our estimates here. \cite{Ueda2014} derive the absorption function in the local Universe using data from the BAT 9-month survey. They find similar fractions for Compton-thick AGN in the same column density range ($\rm N_H=10^{24-25}~cm^{-2}$). In general, it appears that a consensus has been reached at least regarding the numbers of Compton-thick AGN with $\rm N_H=10^{24-25}~cm^{-2}$ in the local Universe \citep[see also][]{Burlon2011,Georgantopoulos2019,Torres-Alba2021}. All the above have been derived on the basis of BAT detections. This is not surprising since the large pass-bands of the \swift mission facilitates the detection of Compton-thick AGN.
 
 Recently, \cite{Boorman2024} constrained the Compton-thick fraction in the local Universe (z$<$0.044), using a sample of 122 AGN, primarily selected to have warm {\it IRAS} colours. By fitting the available X-ray spectra, they estimate a Compton-thick fraction of 35$\pm$6 \%. 
 In a similar,  \cite{Akylas2024} used a sample of {\it WISE}
  selected AGN up to redshifts of $z=0.02$. The estimated Compton-thick fraction is 0.25$\pm$0.05 \%.

Beyond the local Universe, the results  based primarily on the modelling of \chandra and \xmm spectra, present a considerable scatter.  A summary of the estimated fraction of Compton-thick AGN in the $\rm N_H=10^{24-25}~cm^{-2}$ column density range is given in Table \ref{fractions}. 
We note that some of these works \citep{Buchner2015,Laloux2023} quote the fraction of Compton-thick AGN in the column density range 
 $\rm N_H=10^{24-26}~cm^{-2}$ despite the fact that the number of detected sources with $\rm >10^{25} ~cm^{-2}$
 is extremely small \citep[see discussion in][]{Buchner2015}. For this reason, it can be safely assumed  that the 
 observed fraction  of Compton-thick AGN in the $\rm N_H=10^{24-25}~cm^{-2}$ column density range 
 is approximately equal to the fraction of Compton-thick sources in the $\rm N_H=10^{24-26}~cm^{-2}$ range.

 It is unclear whether the fraction of Compton-thick AGN increases with redshift. \citep{Buchner2015} find that the fraction is consistent with being constant while  \cite{Lanzuisi2017} find a steep increase at high redshifts.
 If indeed the fraction increases with redshift, 
 this may mark a genuine evolution of the obscuring medium in Compton-thick AGN with cosmic time. For example, it has been proposed that the excess obscuration at higher redshifts is associated with the host galaxy of the AGN \cite[e.g.][]{Gilli2022}. Alternatively, the excess number  of Compton-thick AGN could be an artefact of the moderate photon statistics at higher redshifts combined with the limited pass-band of the \chandra and \xmm missions. Interestingly, \cite{Laloux2023} analyse the X-ray spectra of AGN in the COSMOS field using a novel method. They derive the 12$\mu m$ luminosity of the AGN component to use it as a prior for the determination of the X-ray obscuration. At  redshifts of $z<0.5$, their method yields results which are compatible with the BAT results in the local Universe. At higher redshifts \cite{Laloux2023}  could derive only upper limits for the Compton-thick fraction.

\subsection{NuSTAR number counts constraints}
Here, we explore the possible constraints that can be posed by the 
serendipitously selected {\it NuSTAR}
Compton-thick AGN. 
\cite{Lansbury2017a} present the sources that have been detected in the 
40-month serendipitous source catalogue covering 13 $\rm deg^2$. 
 Making use of this catalogue, \cite{Lansbury2017b}  select the candidate Compton-thick AGN by applying a hardness ratio criterion, hard to soft band ratio $BR_{Nu}>1.7$ using the 3-8 keV and 8-24 keV bands. This hardness ratio has been chosen based on the spectrum of a Compton-thick AGN assuming the model of \cite{Balocovic2014}.
 They find four candidate sources at small redshifts within  $z<0.07$. Subsequent \nustar spectral analysis \citep{Lansbury2017b}  appears to be consistent with Compton-thick absorption.
Recently, \cite{Greenwell2024} presented the new serendipitous 
\nustar catalogue covering 40 $\rm deg^2$. They find three additional Compton-thick AGN by applying the 
same hardness ratio criterion, $BR_{Nu}>1.7$, within 
the same redshift range. Out of these seven sources only four are within the range of our derived 
luminosity function. 
We use these candidate Compton-thick sources to derive the number count distribution in the 8-24 keV band 
in order to compare with our Compton-thick
 luminosity function. In Fig. \ref{NumberCounts} (left) we compare the \nustar 
and BAT number number counts within $z=0.05$ with the predictions of our luminosity function. In the same figure (right), we compare the \nustar number counts within $z=0.07$ with the predictions of our luminosity function extrapolated to $z=0.07$, in order to exploit the better number statistics. We assign weights for each source in the \nustar number counts.
 For sources with X-ray spectral fits in \cite{Lansbury2017b}
  the weight is calculated taking the full $\rm N_H$ distribution into account. For the remaining sources we use the error on the 
  $BR_{Nu}$ hardness ratio as the weight. The errors 
  on the BAT and \nustar number counts are estimated 
  following \cite{Mateos2008}.

In the same plot we give the predictions of the \cite{Ananna2019} X-ray background synthesis model. 
It appears that our luminosity function lies well below the \nustar number counts  in the $z<0.07$ redshift range. The luminosity function 
of \cite{Ananna2019} derived from their X-ray background synthesis 
model, which actually includes the constraints of the \nustar 
Compton-thick number counts, is actually much closer. 
Their model includes a number of reflection dominated Compton-thick $\rm N_H=10^{25-26}~cm^{-2}$ AGN equal to the number of the 
   transmission-dominated Compton-thick AGN. Taken a face value,
  this result could imply that the vast majority of the Compton-thick  population are associated with reflection dominated AGN which remain undetected by BAT owing to its limited sensitivity. 
  However, caution has to be exercised in the 
  interpretation of the \nustar number counts.
  \cite{Akylas2019} demonstrated that the \nustar number counts 
 of the full X-ray source population  in the 8-24 keV and the 3-8 keV bands are incompatible with the BAT and the \chandra number counts respectively. In particular, there is an upturn  in the 
 8-24 keV \nustar number counts at fluxes of a few $\rm 10^{-14}~erg~cm^{-2}~s^{-1}$. This could be attributed to a 
 very strong Eddington bias at these faint fluxes \cite{Civano2015}.

\subsection{A missing population of Compton-thick AGN?}
Additional constraints on the number of Compton-thick AGN can be provided by the X-ray background synthesis models.
As we have explained above, many of these models assume the existence of large numbers of extremely obscured Compton-thick sources with $\rm N_H>10^{25}~cm^{-2}$. However, given the uncertainties of the X-ray background spectrum at high energies, around 30 keV, it is by no means certain that the addition of these reflection-dominated sources is required to fit the background \cite[e.g][]{Akylas2012}. Actually, \cite{comastri2015} argue that a large number of heavily buried Compton-thick AGN could be present without violating the X-ray and IR  background constraints. NGC 4418 is usually considered as the prototype for this population \citep[e.g.][]{Sakamoto2010}. One way to compile large samples of these extreme sources would be to study in X-rays volume-limited samples of nearby AGN detected in either the optical or the IR bands. 
\cite{Akylas2009} have examined the \xmm spectra of the Seyfert-2 galaxies in the \cite{Ho1997} sample of nearby galaxies. They find a Compton-thick AGN fraction compatible with the findings in this work. More recently, \cite{Asmus2020} have compiled the most comprehensive sample of candidate AGN in the nearby Universe, z$<$100 Mpc. Detailed studies of these sources with \nustar provide a significant advance in the study of the 
less luminous local Compton-thick AGN \citep{Akylas2024}.

\section{Summary}
We compile a new sample of bonafide Compton-thick AGN based on the initial selection of candidates from the 70-month BAT survey. Then, we confirm that these are Compton-thick sources using the column densities derived by \nustar spectral analysis. Our final sample  consists of 44 sources up to a redshift of $z=0.05$ with intrinsic luminosities as faint as $L_X\approx3\times10^{41}$ $\rm erg~s^{-1}$ in the 10-50 keV band. All these sources have column densities in the range $\rm N_H=10^{24-25}~cm^{-2}$. Our primary goal is to derive a robust X-ray luminosity function for Compton-thick AGN in the local Universe and based on this to securely estimate the fraction of Compton-thick AGN. The derivation of the luminosity function follows a Bayesian methodology where the errors on the column densities and the luminosities are fully taken into account. Our results can be summarised as follows:

\begin{itemize}
    \item The luminosity function is described with a double power-law where the faint and bright end slope are $\gamma_1=0.01^{+0.51}_{-0.74}$ and $\gamma_2=1.72^{+0.40}_{-0.37}$, respectively. The break of the luminosity function is $\rm logL_*[erg~s^{-1}]=42.9^{+0.44}_{-0.43}$. The flat slope of the faint end of the luminosity function rather argues against a numerous population of faint Compton-thick AGN
    
    \item The fraction of Compton-thick AGN relative to the total AGN population is 24$\pm5$\%. This is estimated using our Compton-thick luminosity function and the Compton-thin luminosity function of \cite{Ueda2014}.
   \end{itemize}

In conclusion, there appears to be a consensus on the number of Compton-thick AGN in the local Universe at least when these are derived from the BAT data.  
At higher redshifts a significant scatter is observed in the 
estimated fraction of Compton-thick AGN. 
If a higher fraction is indeed confirmed by future studies 
such as those that will be performed with the {\it ATHENA} mission,
this would suggest a strong evolution of the AGN Compton-thick population with cosmic time.

\begin{acknowledgements}
 We thank the anonymous referee for very constructive comments. 
 We acknowledge financial support by the European Union’s Horizon 2020 programme “XMM2ATHENA” under grant agreement No 101004168. The research leading to these results has also received funding from the European Union’s Horizon 2020 Programme under the AHEAD2020 project (grant agreement n. 871158).
\end{acknowledgements}

 
\bibliography{ref}{}
\bibliographystyle{aa}
\appendix

\onecolumn

\end{document}